\documentclass[conference]{IEEEtran}
\IEEEoverridecommandlockouts
\usepackage{kotex}
\usepackage{parskip}
\usepackage{cite}
\usepackage{amsmath,amssymb,amsfonts}
\usepackage{multirow}
\usepackage{algorithmic}
\usepackage{graphicx}
\usepackage{textcomp}
\usepackage{xcolor}
\def\BibTeX{{\rm B\kern-.05em{\sc i\kern-.025em b}\kern-.08em
    T\kern-.1667em\lower.7ex\hbox{E}\kern-.125emX}}
\begin{document}

\title{Hidden Costs for Inference with Deep Network \\ on Embedded System Devices
\thanks{This research was supported by Basic Science Research Program through the National Research Foundation of Korea (NRF) funded by the Ministry of Education(RS-2024-00358953)}
}

\author{\IEEEauthorblockN{Chankyu Lee}
\IEEEauthorblockA{\textit{Department of Electronics Engineering} \\
\textit{Gangneung-Wonju National University}\\
Gangneung, South Korea \\
cksrb0710@gwnu.ac.kr}
\and
\IEEEauthorblockN{Woohyun Choi}
\IEEEauthorblockA{\textit{Department of Advanced AI} \\
\textit{LG Electronics}\\
Seoul, South Korea \\
settler12@gmail.com}
\and
\IEEEauthorblockN{Sangwook Park*}
\IEEEauthorblockA{\textit{Department of Electronics Engineering} \\
\textit{Gangneung-Wonju National University}\\
Gangneung, South Korea \\
spark2@gwnu.ac.kr}
{\footnotesize \textsuperscript{*}corresponding author}
}

\maketitle

\begin{abstract}
This study evaluates the inference performance of various deep learning models under an embedded system environment. In previous works, Multiply-Accumulate operation is typically used to measure computational load of a deep model. According to this study, however, this metric has a limitation to estimate inference time on embedded devices. This paper poses the question of what aspects are overlooked when expressed in terms of Multiply-Accumulate operations. In experiments, an image classification task is performed on an embedded system device using the CIFAR-100 dataset to compare and analyze the inference times of ten deep models with the theoretically calculated Multiply-Accumulate operations for each model. The results highlight the importance of considering additional computations between tensors when optimizing deep learning models for real-time performing in embedded systems.
\end{abstract}


\section{Introduction}
Consumers encounter a variety of deep learning-based services in their daily lives, ranging from mobile devices and home appliances to automobiles. These services are offered by manufacturers who continually strive to develop advanced deep learning techniques to enhance accuracy and ensure customer satisfaction. Since deep learning gained prominence, accuracy has become the primary goal in technology development. Over time, increasingly large and complex models have been introduced, integrating various blocks such as attention mechanisms, residual connections, and the compression and expansion of feature dimensions ~\cite{vaswani2017attention}~\cite{he2016deep}~\cite{Howard2017}. However, these models require substantial resources in terms of computation, memory storage, and power consumption during both training and inference phases. Acknowledging this challenge, recent advancements in deep models are being pursued with careful consideration of not only accuracy but also the model's size (i.e. the number of parameters) and computational efficiency ~\cite{tan2019efficientnet}.

Generally, there is a trade-off between reducing the number of model parameters and improving model accuracy. Therefore, when deploying deep models on devices with limited memory resources, the objective is to minimize the number of parameters while maintaining accuracy. Intuitively, one would expect that as model size decreases, the computation during inference phase would also decrease. However, this expectation can be misleading depending on the model architecture. An example of handwritten digit classification using the MNIST dataset under an embedded system is summarized in Table~\ref{tab:mnist}. In this sample test, three models are evaluated. Model A is a standard model composed of double stacking regular convolution layers with large-sized kernels and a single fully-connected layer. In case of Model B, each of the convolutions in Model A is replaced with double stacking regular convolution with small-sized kernels to reduce the number of model parameters but maintain the size of receptive field~\cite{Simonyan2014}. And, Model C is designed by replacing each convolution of Model A with separable convolution modules, depthwise- and pointwise-convolution~\cite{Howard2017}. Note that the accuracy is a little decreased due to exclusion of batch normalization in separable convolution module to ensure a fair comparison of computation times. Interestingly, this example demonstrates that even with fewer model parameters and MAC operations, computation times can still be high. This phenomenon presents challenges in accurately assessing the resource requirements of systems implementing deep learning-based services.


\begin{table}[ht]
\caption{A Sample Test of Handwritten Digit Classification}
\begin{center}
\renewcommand{\arraystretch}{1.3} 
\begin{tabular}[\columnwidth]{|c|c|c|c|c|}
\hline
  &  & \textbf{Model A} & \textbf{Model B} & \textbf{Model C} \\
\hline
\hline
\multirow{10}{*}{Archi-} & \multirow{2}{*}{Conv1} & \multirow{2}{*}{5x5, 10-ch} & 3x3, 5-ch  & 5x5, dw, 1-ch \\
\multirow{10}{*}{tecure}&      &            & 3x3, 10-ch & 1x1, 10-ch \\
\cline{2-5} 
 &  Act.  &  \multicolumn{3}{c|}{ReLU} \\
\cline{2-5} 
 &  Pool  &  \multicolumn{3}{c|}{MaxPooling stride=2} \\
\cline{2-5} 
  & \multirow{2}{*}{Conv2} & \multirow{2}{*}{5x5, 20-ch} & 3x3, 15-ch & 5x5, dw, 10-ch \\
  &       &            & 3x3, 20-ch & 1x1, 20-ch \\
\cline{2-5} 
 &  Act.  &  \multicolumn{3}{c|}{ReLU} \\
\cline{2-5} 
 &  Pool  &  \multicolumn{3}{c|}{MaxPooling stride=2} \\
\cline{2-5} 
  &  fc  & \multicolumn{3}{c|}{Linear(980, 10)} \\
\cline{2-5} 
  &  Out  & \multicolumn{3}{c|}{Softmax} \\
\hline
\hline
\multirow{2}{*}{Comp-} & \# parameters & 15,090 & 12,650 & 10,336 \\
\cline{2-5} 
\multirow{2}{*}{lexity} & MACs & 1,185,800 & 616,616 & 125,440 \\
 \cline{2-5} 
  & inf. time$^{\mathrm{a}}$ & 3.298 ms & 4.003 ms & 6.033 ms \\
\hline
\multicolumn{5}{c}{$^{\mathrm{a}}$ Average time of inference for 100-samples on Embedded system.}
\end{tabular}
\label{tab:mnist}
\end{center}
\end{table}


In previous studies, both the number of model parameters and Multiply-Accumulates (MACs) are used to assess model complexity and computational efficiency~\cite{Howard2017}. On the other hand, as demonstrated in Table~\ref{tab:mnist}, there is a discrepancy between these metrics and computation time. This inconsistency complicates the prediction of resource requirements for embedded systems. This paper provides inference times of 10-representative models for image classification, and explores the relationship between model components and computation time. The experiments focus on image classification tasks and evaluate 10-representative models based on the CIFAR100 dataset. Computation times are measured on embedded system hardware featuring a 1.8GHz processor and 8GB of RAM, and the results are compared with each model's accuracy, the number of parameters, and computational load to propose efficient design strategies.

\section{Materials and Methods}
\subsection{Database \& Hardware specification}
This paper focuses on the image classification task to investigate inference time on embedded device with a processor specified as Quad-core 64bit@1.5GHz with 8GB of RAM. CIFAR-100 dataset is used to train and evaluate each of deep models in model accuracy and inference time.
\subsection{Models}
This study performs image classification task with 10-models: \textit{VGG16}, \textit{InceptionV3}, \textit{InceptionV4}, \textit{ResNet50}, \textit{SeResNet50}, \textit{Xception}, \textit{MobileNet}, \textit{MobileNetV2}, \textit{ShuffleNet}, and \textit{ShuffleNetV2}. In a nutshell, \textit{VGG16} consists of 16-layers using 3x3 convolutional layers along with 2x2 max pooling layers. This model suggests multiple convolution with small sized kernels to extend receptive field with fewer model parameters~\cite{Simonyan2014}. \textit{Inception} has a structure that employs convolution filters of various sizes in parallel to extract feature effectively. In here, two versions of \textit{Inception}, consist of 94-layers for version 3 and 149-layers for version 4 respectively, are considered~\cite{szegedy2016rethinking,szegedy2017inception}. \textit{ResNet50} is composed of 50-residual blocks, which use skip connection over upper-layers for stable learning~\cite{he2016deep}. \textit{SeResNet50} incorporates Squeeze-and-Excitation blocks, which enhance performance by weighting important information across channels into \textit{ResNet50}~\cite{hu2018squeeze}. \textit{Xception} is an extension of the \textit{Inception} model that achieves high performance with fewer parameters by using depthwise separable convolutions~\cite{chollet2017xception}. \textit{MobileNet} is a lightweight model designed for inference on mobile devices, utilizing depthwise separable convolutions~\cite{Howard2017}. \textit{MobileNetV2} is a subsequent model that further enhances the performance of the \textit{MobileNet} by employing an Inverted Residual structure and bottlenecks \cite{sandler2018mobilenetv2}. \textit{ShuffleNet} is an extremely efficient version of \textit{MobileNet} by applying channel shuffling ~\cite{zhang2018shufflenet}. And, \textit{ShuffleNetV2} provides an advanced version of \textit{ShuffleNet} in the balance between efficiency and accuracy in lightweight network~\cite{ma2018shufflenet}.

The training and inference for the ten models described above are implemented in a Python environment using the PyTorch library.

\begin{figure}[ht]
\centerline{\includegraphics[width=0.8\columnwidth]{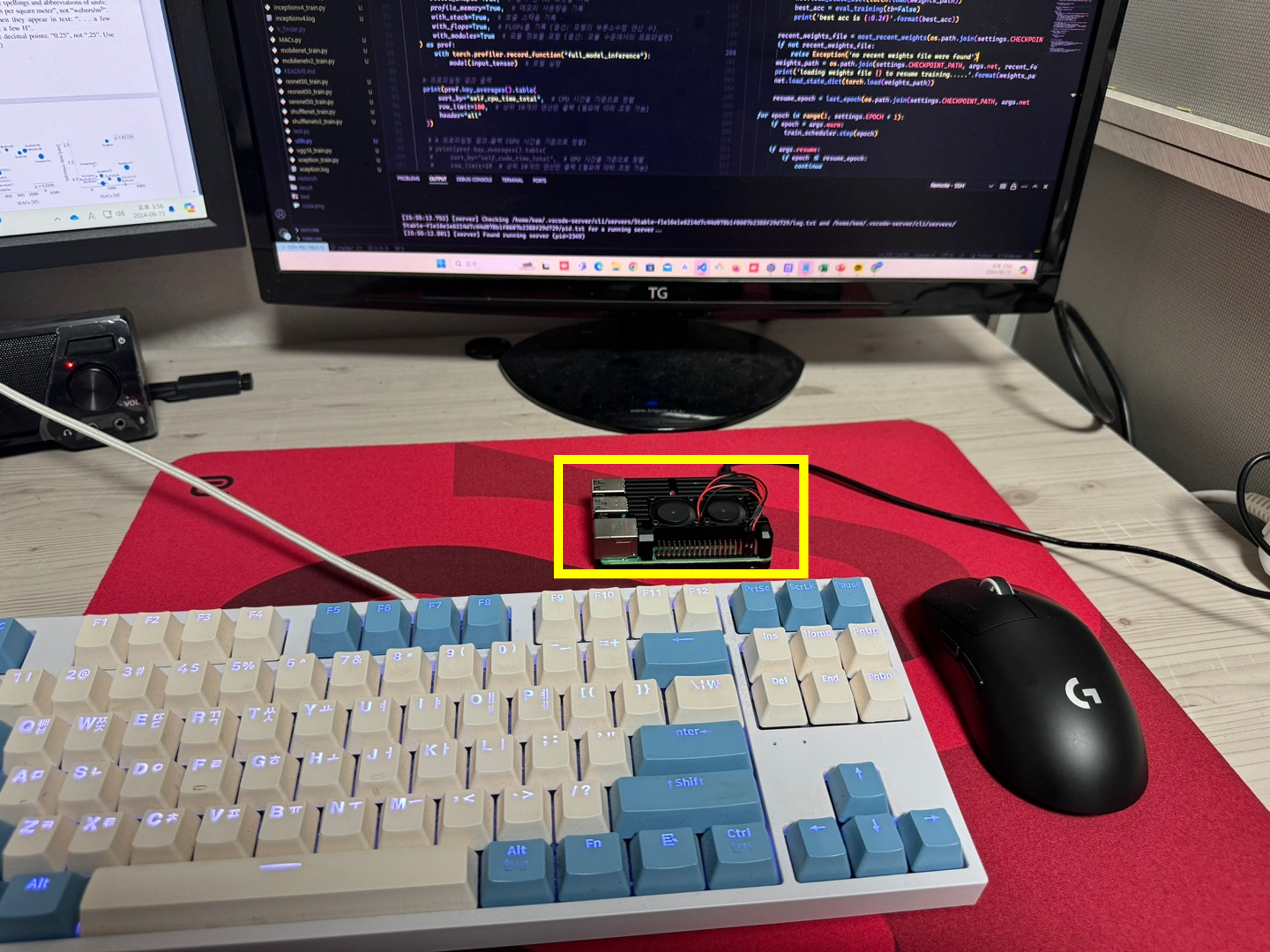}}
\caption{Experiment environment performed on embedded system}
\label{fig}
\end{figure}

\subsection{Survey of inference time and MAC}
In this study, two types of library are used to measure inference time and MACs. With \textit{torch.profile}, inference time is measured for each model on embedded system device while \textit{thop.profile} is applied to calculation of MACs using an input. The calculation method is as follows~\cite{Howard2017}:

\textbf{Convolutional Layers:}
\begin{equation}
\text{MACs} = 2 \times (C_{\text{in}} \times H_{\text{out}} \times W_{\text{out}} \times K_H \times K_W \times C_{\text{out}}) \label{eq1}
\end{equation}

\textbf{Fully Connected Layers:}
\begin{equation}
\text{MACs} = 2 \times (N_{\text{in}} \times N_{\text{out}}) \label{eq2}
\end{equation}

where, \(C_{\text{in}}\) represents the number of input channels, while \(H_{\text{out}}\) and \(W_{\text{out}}\) denote the height and width of the output feature map, respectively. \(K_H\) and \(K_W\) correspond to the height and width of the convolutional kernel, and \(C_{\text{out}}\) refers to the number of output channels. For fully connected layers, \(N_{\text{in}}\) represents the number of input neurons, and \(N_{\text{out}}\) represents the number of output neurons.

\begin{figure*}[ht]
    \centering
    \includegraphics[width=\textwidth, height=4.3cm]{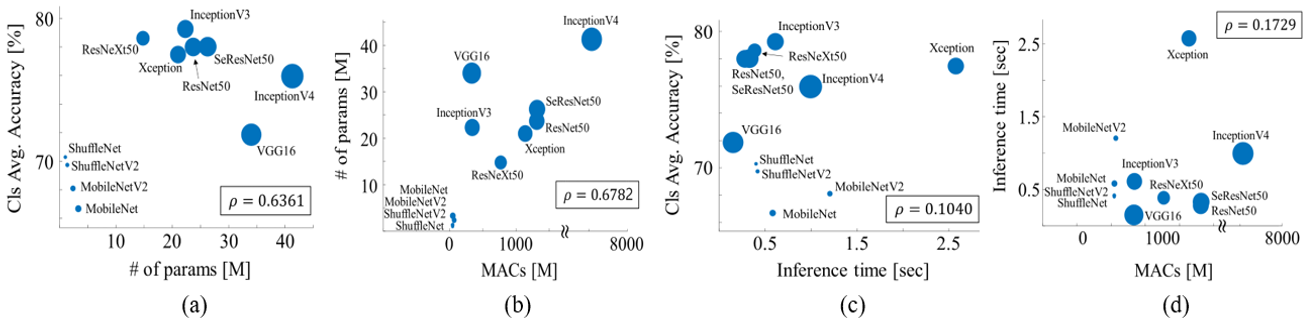}
    \caption{Distributions of model accuracy and computation speed for 10-deep models along with Pearson correlation coefficient: (a) Accuracy vs size, (b) Size vs Computations, (c) Accuracy vs time, (d) Time vs computations}
    \label{fig:inference_time}
\end{figure*}


\begin{figure}[ht]
    \centering
    \hspace{-1.3cm} 
    \includegraphics[width=0.8\columnwidth]{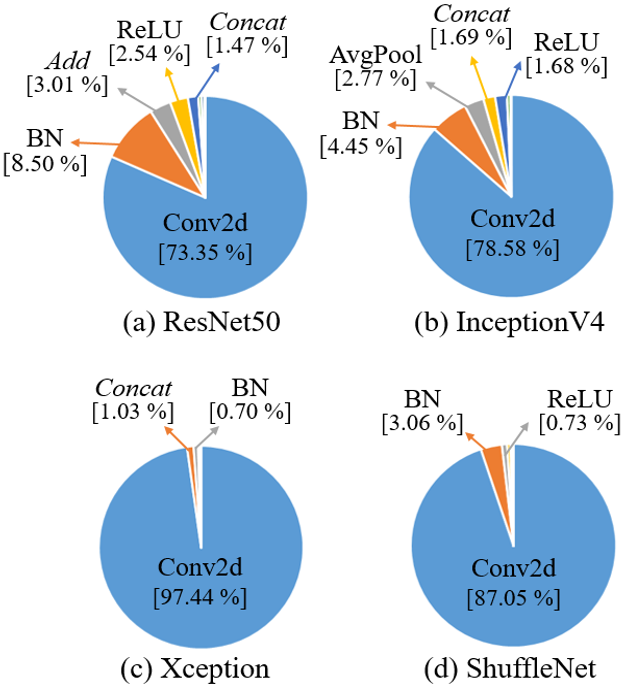}
    \caption{The ratio of computation components during model inference.}
    \label{fig:computation_ratio}
\end{figure}

\section{Results}
The number of model parameters, MACs, class averaging accuracy, and inference time are summarized in Fig~\ref{fig:inference_time}, along with the Pearson correlation coefficient, $\rho$. Note that the size of circles represents the model size. The relationships between model accuracy and model size, as well as between model size and computations, are consistent with the intuitions of the deep learning paradigm. It is understandable that there is no correlation between model accuracy and inference time, since increased computations do not always translate to performance enhancement. However, the absence of correlation between inference time and computations warrants further investigation.

To conduct a more detailed analysis, four-models such as ResNet50, InceptionV4, Xception, and ShuffleNet, which differ significantly in their MACs and inference time, are selected. The inference time is then factorized into individual operations as summarized in Fig.~\ref{fig:computation_ratio}. All four-models execute Conv2D operations for more than 70\% of their total inference time followed by Batch Normalization (BN), ReLU activation, concatenation (Concat), and others. It is important to note that both \textit{Concat} and \textit{Add} operations are performed between tensors rather than model components, and these operations are excluded from the MAC calculations. 

To perform a more detailed analysis, computing time of functions invoked during Conv2d operations are summarized in Table~\ref{tab:conv2d}. The Conv2d operation encompasses various functions, including convolution, slice, narrow, empty, and view. Specifically, slice and narrow are used to extract or exclude specific dimensions of data from multi-dimensional tensor. The empty function allocates the memory storage required for the convolution operation, while view modifies the shape of the tensor without additional memory storage. All these operations are tensor-specific and are excluded in the MAC calculations.

In terms of computation time, the convolution operation accounts for over 97\% of the total operation time for the ResNet50 and InceptionV4, whereas the others exhibit relatively lower percentages for the convolution operation. Notably, in the case of Xception, only about 50\% of the computations occur within the convolution layer. This discrepancy highlights the difference between the MAC of the Xception model and its actual computation time. For high-performance devices such as servers, such tensor operations are typically performed without significant delays (\textit{See, Table~\ref{tab:append}}). However, it is crucial to recognize that on devices with limited hardware resources, the time delays or power consumption associated with such tensor operations can have a substantial impact.


\begin{table}[htbp]
    \caption{Computation time of each component in Conv2d operation [ms]}
    \begin{center}
        \renewcommand{\arraystretch}{1.3} 
        \begin{tabular}{|c|c|c|c|c|c|}
        \hline
         &  Ops. & \textbf{ResNet50} & \textbf{InceptionV4} & \textbf{Xception} & \textbf{ShuffleNet} \\
        \hline
         & convolution & 200.67 & 766.48 & 1269 & 190.65 \\
         & \textit{slice}$^*$  & 0 & 0.99 & 389.73 & 53.53 \\
         & \textit{narrow}$^*$ & 0 & 1.38 & 602.67 & 83.70 \\
         & \textit{empty}$^*$ & 3.05 & 10.17 & 182.43 & 15.60 \\
         & \textit{view}$^*$ & 0.66 & 1.53 & 47.24 & 5.27 \\
         & etc & 2.17 & 1.44 & 18.9 & 0.78 \\
         \hline
         \multicolumn{2}{|c|}{total}  & 206.55 & 781.99 & 2510 & 349.53 \\
        \hline
        \multicolumn{6}{c}{$^*$ tensor operation} \\
        \end{tabular}
        \label{tab:conv2d}
    \end{center}
\end{table}

\section{Conclusion}
This study examines the relationship between inference time and computational complexity in various deep learning models, revealing the limitations of MACs (Multiply-Accumulate Operations) as a performance metric. While MACs focus primarily on operations of model components such as convolutions, matrix multiplication, activation function, they often fail to account for additional computational costs, such as slicing a tensor and memory allocation, which can significantly impact actual inference time. These auxiliary operations are not included in the MACs metrics but can still influence considerable delays in computation. For instance in Xception, despite having relatively fewer MACs than InceptionV4, the higher proportion of operations like slicing can result in longer inference time. This inefficiency becomes even more pronounced in resource-constrained embedded systems, where the impact on performance is more significant. Consequently, when designing deep learning models for such systems, it is insufficient to consider only layer configuration; other operations that contribute to computational overhead must also be accounted for to achieve optimal performance on embedded system devices.

\begin{table}[]
    \caption{Appendix A. Computation components of Conv2d operation in different devices: performed either on a 24-core CPU@3.0 GHz, RAM 125 GB or GPU@8192 processing units, 24 GB of GPU memory.}
    \begin{center}
        \renewcommand{\arraystretch}{1.3} 
        \setlength{\tabcolsep}{5pt} 
        \begin{tabular}{|c|c|c|c|c|c|}
        \hline
        & [ms]  & \multicolumn{2}{|c|}{InceptionV4} & \multicolumn{2}{|c|}{Xception} \\
        \hline
         &  Ops. & \textbf{local PC} & \textbf{local PC-gpu} & \textbf{local PC} & \textbf{local PC-gpu} \\
        \hline
         & convolution &34.17 &7.4 &12.83 &2.23 \\
         & \textit{cpu: slice}$^*$ &0.07 &0.81 &0 &0.40 \\
         & \textit{cpu: narrow}$^*$ &0.13 &0.06 &0 &0.005 \\
         & \textit{empty}$^*$ &0.24 &0.82 &0.08 &0.33 \\
         & \textit{view}$^*$ &0.007 &0.01 &0.009 &0.007 \\
         & etc &1.25 &0.3 &0.93 &0.22 \\
         \hline
         \multicolumn{2}{|c|}{total}  &36.14 &9.4 &13.85 &3.19  \\
        \hline
        \multicolumn{6}{c}{$^*$ tensor operation} \\
        \end{tabular}
        \label{tab:append}
    \end{center}
\end{table}

\bibliographystyle{IEEEbib}
\bibliography{references.bib}

\vspace{12pt}
\end{document}